\documentclass[prd,nofootinbib,showpacs,showkeys]{revtex4}
\usepackage{indentfirst}
\usepackage{latexsym}
\usepackage{graphicx}

\begin{document}
\title{The Application of Bayesian Technique for
Particle Identification}
\author{Ding Tian}
\email{tianding@cugb.edu.cn}
\affiliation{Teaching and Research
Section of Physics, School of Materials Science and Technology,
China University of Geosciences(Beijing), Beijing 100083, P.R.
China}

\begin{abstract}
The PID problem in high energy physics experiments
is analysed with Bayesian technique. The corresponding applicable
method is presented.
\end{abstract}
\pacs{02.50.Cw, 07.05Kf, 29.90.+r}
\keywords{Particle identification; Bayesian technique}
\maketitle

\section{Introduction}
Particle identification(PID) is important in high energy physics
experiments, and it mainly refers to  charged particles. Different
techniques\cite{haykin,lonnblad, hultqvist,pallavicini}are used to
study this problem. In this paper, the problem is analysed with
Bayes' theorem of probability theory. It is well known that the best
classification methods are based on Bayesian techniques if all the
probability distributions are known\cite{pallavicini}. However, from
a literature survey, it appears that how to use Bayesian technique
in PID problem has not yet been thoroughly
investigated\cite{aguilar,taka,kuo}.

Different detectors use different variables to do PID, such as
TOF $t$ from TOF detector, $dE/dx$ from wire chamber, deposited
energy $E$ from shower counter, Cherenkov radiation emission
angle $\theta$ from RICH counter, the deposited
energy $W$ or transition radiation(TR) photon hits $N$ from TR
detector, etc. For different particles with same momentum, the
random variables($t$, $dE/dx$, $E$, $\theta$, $W$, $N$ etc)
\footnote{In this paper, random variable and its value are denoted
with same symbol.} may have different distributions which can be
used for PID, therefore in this paper we call
the random variables PID variables.
Sometimes more than one PID variables which have different character
can be obtained from one detector, such as in shower counter, both
the deposited energy $E$ of a shower and one or two variables which describe
the shape of the shower can be used for electron/hadron separation.

For an unknown charged particle, its momentum is usually known(e.g.,
given by drift chamber).
Therefore, all the calculations of probabilities in this paper are under the
condition that the particle's momentum vector is known and indicated with
$p$, $\theta$ and $\phi$ which are magnitude, polar and azimuthal angles of
the particle's momentum vector respectively.

The paper is organized as follows: In section two, the PID problem is
analysed with Bayesian technique when there is only one PID variable
(use TOF $t$ as example)is obtained for an unknown charged
particle. In section three, similar analysis is done
when two and more PID variables(use TOF $t$ and the deposited energy
$E$ in shower counter as example)are available.
Section four is the conlusions.

\section{Case for one PID variable}

For a fixed momentum denoted with the parameters $p$, $\theta$ and $\phi$,
$P(i)(i=1,2,3,4,5)$ are used to represent the appearing probabilities
of particle
$\mathbf{e^+,\mu^+,\pi^+,K^+,p^+}$(or $\mathbf{e^-,\mu^-,\pi^-,K^-,p^-}$)
\footnote{The unknown particle's charge is known.}
respectively, because five kinds of
particles can have the same parameters' values $p$, $\theta$, $\phi$.
Here and below, $i$ and $j$ are used to represent one particle in
$\mathbf{e^+,\mu^+,\pi^+,K^+,p^+}$ or $\mathbf{e^-,\mu^-,\pi^-,K^-,p^-}$.
And because only five kinds of particle can be the unknown particle, the
appearing probabilities should be normalized to unit for the fixed momentum:
\begin{equation}
\sum_{i=1}^5 P(i)=1.
\end{equation}
if some kind particle does not appear, the corresponding $P(i)=0$; and if
the number of charged particle kinds is larger than five(e.g., cosmic rays
or particles from nuclear reaction), the sum terms will exceed five.

When there is only one PID variable(e.g. TOF $t$), what we know is the
TOF $t$ of the unknown charged particle and the conditional probability
$P(t|i)$ which is the probability of TOF $t$, given that the
unknown charged particle is $i$.
\footnote{Because TOF $t$ has continuous distribution, the values of
$P(t|i)$ are all infinitesimals. } From the point of view of
probability theory,
only the probability that the unknown charged particle is $i$
can be determined. Then the PID
problem can be written as follows: \\ \hspace*{0.7cm}Given
the momentum of the unknown charged particle and $P(t|i)$,
calculate $P(i|t)$, \\ where
$P(i|t)$ is the conditional probability that the unknown charged particle
is $i$, given that the TOF of
the unknown charged particle is $t$.
In the light of the definition of conditional
probability and Bayes' theorem, we have
\begin{eqnarray} \label{eq1}
P(i|t) & = & \frac{P(t|i)P(i)}{P(t)}
   = \frac{P(t|i)P(i)}{\sum\limits_{j=1}^{5}P(t|j)P(j)}\nonumber \\
 & = & \frac{f_i(t)dt\cdot P(i)}{\sum\limits_{j=1}^{5} f_j(t)dt\cdot P(j)}
   = \frac{f_i(t)P(i)}{\sum\limits_{j=1}^{5} f_j(t)P(j)}
\end{eqnarray}
where $P(j)$ is the appearing probability
of the charged particle $j$, $P(t)$ is the
probability that TOF $t$ occurs, and
$f_j(t)$ is the probability density function(p.d.f.) of variable $t$
for the charged particle $j$.
The denominator in equation(\ref{eq1}) is the normalizing constant which
only makes $P(i|t)$ have the probability meaning. The probability
$P(i|t)$ is proportional to $f_i(t)P(i)$ in which $f_i(t)$ is determined by
the detector, while $P(i)$ has no concern with any detector.
The p.d.f. for TOF $t$ is usually a Gaussian
distribution, i.e.
\begin{equation}\label{eq2}
f_i(t)=\frac{1}{\sqrt{2\pi}\sigma_i}\exp
[-\frac{(t-t_{i0})^2}{2\sigma_i^2}]
\end{equation}
where $\sigma_i$ is the
resolution of TOF for the charged particle $i$,
$t_{i0}$ is the expected value of TOF for the charged particle $i$.
The general result for above pattern recognition can be easily
found\cite{duda}.

According to the physical meaning of $P(i|t)$,
after five values $P(i|t)$ have been calculated,
the reasonable hypothesis for the unknown charged particle is $i$
which makes $P(i|t)$ the largest in the five values.
For any other PID variable $X$, if all p.d.f.s of TOF $t$ in
equation(\ref{eq1}) are replaced with the corresponding p.d.f.s of
variable $X$, equation(\ref{eq1}) can be used for PID variable $X$.
But the PID variable $X$ may not have a Gaussian distribution for
every $i$ as TOF $t$ has in equation(\ref{eq2}). For example,
the deposited energy $E$ of a fixed momentum electron in EM shower
counter has Gaussian distribution, while for $\pi$,$K$,
the deposited energy usually has not.

In equation(\ref{eq1}), the p.d.f.s $f_i(t)$ (or $P(t|i)$) can be
obtained from calibration of the detector.
Thus the appearing probability(or prior probability)$P(i)$ is the
only unknown quantity. And it is $P(i)$ that makes PID problem
complicatedly because $P(i)$ varies with studied final states.
Here, we give some remarks on $P(i)$.
\begin{enumerate}
 \item  $P(i)$ is the appearing probability of the charged particle
        $i$ for studied
        final state and the momentum vector($p$, $\theta$, $\phi$).
        This means different final states have different $P(i)$, while
        different cuts(e.g. charged track number) in analysis
        result in different final states.
        For example, if all events of $J/\psi$ decay are considered,
        we get a set of $P(i)$ for the momentum vector
        ($p$, $\theta$, $\phi$);
        for the same momentum vector ($p$, $\theta$, $\phi$),
        if only those four-charged-tracks events
        from $J/\psi$ decay are considered,
        we will obtain another set of $P(i)$.
        But why do we need a second set of $P(i)$?
        In fact, the second set of $P(i)$ can be used to
        enhance the efficiency of PID
        when we select the events which only have four charged
        hadron tracks. In the four-charged-tracks events of $J/\psi$ decay,
        the appearing probabilities of leptons ($\mathbf{e, \mu}$) are by far
        less than that of hadrons($\mathbf{\pi,K, p}$).
        If the second set of $P(i)$ is used to select events,
        the affection of leptons will be reduced greatly.
        Therefore, analyses which use corresponding $P(i)$
        will have better event selecting.
        After a series of cuts are used to obtain $P(i)$,
        the correct use of $P(i)$ is that the cuts used in the
        event selecting should not be looser than those cuts used in
        obtainning
        $P(i)$, because the $P(i)$ can not be used to select the events which
        do not belong to the corresponding final state.
        Once $P(i)$ has been figured out, it is not necessary to
        to change it
        when a new analysis is performed so long as the conditions
        which determine $P(i)$ do not change.
 \item  $P(i)$ can be obtained from M.C. process. But a more reliable
        way of obtainning $P(i)$ is recurrence approach in real data.
        M.C. results or theoretical values(if any)
        can be used as initial values.
 \item  If the difference between $P(i)$ is not large,
        PID will mainly rely on the inherent PID capability of detector,
        i.e., p.d.f.s of PID variable(e.g., $f_i(t)$ in equation(\ref{eq1})).
        And if the difference between $\sigma_i$ is neglected,
        we derive the conventional PID method(for TOF detector)
        which is only the
        contribution of exponential part(the weight of the unknown
        charged particle
        to be particle $i$) in equation(\ref{eq2}):
        \begin{equation}
        W_i=\exp[-\frac{(t-t_{i0})^2}{2\sigma_i^2}]
        \end{equation}
    However, the difference between $P(i)$ can not be neglected at will.
    For example, in the final states of $J/\psi$ decay,
    difference between the appearing probabilities of $\pi^{\pm}$, $K^{\pm}$
    varies with momentum from several to ten times\cite{danju}.
    So it is valuable and more accurate to
    consider the effect of $P(i)$ when large difference between $P(i)$
    exists. For example, if the weights of an unknown particle to
    $\pi$, $K$ are equal, i.e., $W_3=W_4$, one may have no idea of
    what the unknown particle is. But arccording to equation(\ref{eq1}),
    the probability which the unknown particle is $\pi$ is several
    to ten times larger than the probability which the unknown particle
    is $K$. Furthermore, if $W_3<W_4$, the particle will be identified
    to be $K$, but $P(3|t)>P(4|t)$ may occur because $P(3)>P(4)$,
    this suggests that the unknown particle is more likely to $\pi$.
    Finally,
    if one does not use $P(i)$, one may have set all $P(i)$
    a same value(equals 0.2)\cite{kuo} which is groundless.

 \item  PID problem will become troublesome if $P(i)$
        depends on three parameters ($p$, $\theta$, $\phi$).
        To reduce the number of the parameters is favourite.
        For final states come from the colliders
        which have equal energy particle and anti-particle colliding,
        it is not difficult to find that $P(i)$ is independent of
        polar angle $\phi$ because of axis symmetry,
        and because there are all kinds of channels in one
        final state(e.g. $J/\psi$ decay or four-charged-tracks
        final state in $J/\psi$ decay),
        $P(i)$ may be independent of azimuthal angle $\theta$.
        Thus, for the final states from most colliders,
        if the cuts of obtainning $P(i)$ are loose enough,
        $P(i)$ may only depend on one parameter $p$,
        the magnitude of momentum vector. In applications,
        the particle's possible momentum region
        can be divided into many small regions
        (e.g. 50MeV/c or less for a region's width). For every region,
        we have five values $P(i)$.
        Then, for an unknown charged particle,
        $P(i|t)$ in which we are interested can be calculated.
\end{enumerate}

Obviously the above procedure has no difficulty of correlations
between particles mentioned in reference\cite{aguilar}.

\section{Case for two and more PID variables}

When there are two PID variables(e.g. TOF $t$ and the deposited energy $E$
in EM shower counter) for one unknown charged particle,
then the PID problem can be written as follows: \\
\hspace*{0.7cm} Given the momentum of the unknown charged particle,
$P(t|i)$ and $P(E|i)$, calculate $P(i|t, E)$,
\\ where $E$ is the measured value of the deposited energy
in shower counter, $P(E|i)$ is the conditional
probability that the deposited energy is $E$ given that
the unknown charged particle is $i$,
and $P(i|t, E)$ is the conditional probability
that the unknown charged particle is $i$,
given that TOF $t$ and the deposited energy $E$ occur simultaneously.
By virtue of the definition of conditional probability, we have again
\begin{eqnarray}
P(i|t, E) & = & \frac{P(i, t, E)}{P(t, E)}
          =\frac{P(t, E|i)P(i)}{P(t, E)}
\end{eqnarray}
where $P(i,t,E)$ is simultaneous occurrence probability of
$i$,$t$ and $E$; $P(t,E)$ is the probability that
TOF $t$ and the deposited energy $E$ occur simultaneously;
$P(t,E|i)$ is the conditonal probability that
TOF $t$ and the deposited energy $E$ occur simultaneously
given that the unknown charged particle is $i$.
Because measurements of TOF $t$ and the deposited energy $E$
are independent, we have
\begin{equation}
P(t,E|i)=P(t|i)P(E|i)
\end{equation}
Here, it should be noted that the situation of variable $E$ is not the
same as that of TOF $t$, the probability that $E=0$ may not be
infinitesimal because of finite sensitivity of the detector,
i.e. the distribution of $E$ is not a pure continuous
distribution, but a mixed one:
\begin{equation}
P(E|i)=\left\{ \begin{array}{lr}
P(E=0|i) & \textrm{if}\ E=0; \\
\left[ 1-P(E=0|i)\right] g_i(E)dE & \textrm{if}\ E>0
\end{array} \right.
\end{equation}
where $g_i(E)$ is the p.d.f. of variable $E$ for the charged particle
$i$ when the deposited energy $E>0$.
If $E=0$ for the unknown charged particle, then
\begin{eqnarray}
P(i|t,E=0) & = & \frac{P(t,E=0|i)P(i)}{P(t,E=0)}
   = \frac{P(t,E=0|i)P(i)}
   {\sum\limits_{j=1}^{5}P(t,E=0|j)P(j)} \nonumber \\
 & = & \frac{f_i(t)P(E=0|i)P(i)}{\sum\limits_{j=1}^{5}f_j(t)P(E=0|j)P(j)}
\end{eqnarray}
If $E>0$ for the unknown charged particle, then
\begin{eqnarray}
P(i|t,E>0) & = & \frac{P(t,E>0|i)P(i)}{P(t,E>0)}
   = \frac{P(t,E>0|i)P(i)}
   {\sum\limits_{j=1}^{5}P(t,E>0|j)P(j)}\nonumber \\
 & = & \frac{P(t|i)P(E>0|i)P(i)}{\sum\limits_{j=1}^{5}P(t|j)P(E>0|j)P(j)}
   =   \frac{f_i(t)dt\cdot [1-P(E=0|i)]g_i(E)dE\cdot P(i)}
        {\sum\limits_{j=1}^{5}
        f_j(t)dt\cdot [1-P(E=0|j)]g_j(E)dE\cdot P(j)}\nonumber \\
 & = & \frac{f_i(t)[1-P(E=0|i)]g_i(E)P(i)}
       {\sum\limits_{j=1}^{5}f_j(t)[1-P(E=0|j)]g_j(E)P(j)}
\end{eqnarray}
Similarly, the reasonable hypothesis for the unknown charged particle is $i$
which makes $P(i|t,E)$ the largest in the five values.

Obviously, it is not difficult
to generalize above calculation to the case of many independent
PID variables. Two PID variables from two different detectors are
usually independent.
Furthermore, the method can be used all the
same when a PID variable has discrete distribution
(e.g. $\mu$-detector hits
probability), and using it is straightforward in this case.

If two PID variables $X$ and $Y$ are correlative, the conditional
probability
\begin{equation}
P(X, Y|i)=f_i(X,Y)dXdY
\end{equation}
where $f_i(X, Y)$ is the joint p.d.f. of PID variables $X$ and $Y$
for particle $i$. Similarly, we have
\begin{equation}\label{eq3}
P(i|X, Y)=\frac{f_i(X,Y)P(i)}{\sum\limits_{j=1}^5 f_j(X,Y)P(j)}
\end{equation}
Since the joint p.d.f. $f_i(X,Y)$ is difficult to obtain, the
above equation(\ref{eq3}) is not very useful.

\section{Conclusions}
By employing Bayes' theorem of probability theory, we have
clarified the usage of all types of PID information.
The corresponding applicable method to PID problem is also proposed.
The method
has some attracting properties. First, the final results
(e.g., $P(i|t)$, $P(i|t,E)$) are probabilities which have definite
physical meaning.
Second, when one PID varibale has no-Gaussian distribution
(e.g. Landau distribution of $dE/dx$), this method can be used
as well.
Finally, the conventional PID method can be derived from it
after some approximation.


\begin{thebibliography}{99}

\bibitem{haykin}S.S.Haykin. Neural Networks: A Comprehensive Foundation.
New York: Maxwell-MacMillan, 1994.

\bibitem{lonnblad}L\"onnblad et al.,
Comp. Phys. Commun., 1992, {\bf 70}: 167-182

\bibitem{hultqvist}K. Hultqvist et al., Nucl. phys., 1994, {\bf B421}: 3-37

\bibitem{pallavicini}M. Pallavicini et al., Nucl. Instr. and Meth.,
1998, {\bf A405}: 133-138

\bibitem{duda}R.O.Duda, P.E.Hart. Pattern recognition and Scene Analysis.
New York: Wiley, 1973.

\bibitem{aguilar}M. Aguilar-Benitez et al., Z.Phys.C,
1991, {\bf 50}: 405-426

\bibitem{taka}N. Takashimizu et al., Nucl. Instr. and Meth.,
2004, {\bf A534}: 162-164

\bibitem{kuo}C.C. Kuo et al., Phys. Lett., 2005, {\bf B621}: 41-55

\bibitem{danju}W. Braunschweig et al., Phys. Lett.,
1976, {\bf B63}: 115-127

\end{thebibliography}
\end{document}